\documentclass{PoS}

\title{Sunrise Mission Highlights}

\ShortTitle{Sunrise Mission Highlights}

\author{\speaker{Tino~L. Riethm\"uller},$^a$ Sami~K. Solanki$^{ab}$ and the \textsc{Sunrise} team\\
\llap{$^a$} Max Planck Institute for Solar System Research (MPS), Justus-von-Liebig-Weg 3, 37077 G\"ottingen, Germany\\
\llap{$^b$} School of Space Research, Kyung Hee University, Yongin, Gyeonggi, 446-701, Republic of Korea\\
E-mail: \email{riethmueller@mps.mpg.de}, \email{solanki@mps.mpg.de}}

\abstract{Solar activity is controlled by the magnetic field, which also causes the
variability of the solar irradiance that in turn is thought to influence the
climate on Earth. The magnetic field manifests itself in the form of structures
of different sizes, starting with sunspots (10-50~Mm) down to
the smallest known magnetic features that often have spatial extents of 100~km or less.
The study of the fine scale structure of the Sun's magnetic field has
been hampered by the limited spatial resolution of the available observations.
This has recently changed thanks to new space and ground-based telescopes.
A significant step forward has been taken by the \textsc{Sunrise} observatory, built around
the largest solar telescope to leave the ground, and containing two science
instruments. \textsc{Sunrise} had two successful long-duration science flights on a
stratospheric balloon in June 2009 (solar activity minimum) and in June 2013
(at a high activity level) and a number of scientific results have been
obtained that have greatly advanced our understanding of solar magnetism,
with data analysis still ongoing. After a brief introduction to the \textsc{Sunrise}
mission, an overview of a selection of these results will be given.}

\FullConference{Frontier Research in Astrophysics\\
                 26 - 31 May, 2014\\
                 Mondello (Palermo), Italy}

\usepackage{epsfig}
\usepackage{graphicx}
\usepackage{longtable}

\begin{document}
\newcommand{\sunrise}{\textsc{Sunrise}}
\newcommand\arcsec{\mbox{$^{\prime\prime}$}}
\newcommand{\carcsec}{$\mbox{.\hspace{-0.5ex}}^{\prime\prime}$}

\section{Introduction}

In order to understand which processes control solar activity, we must know how the Sun's magnetic
field interacts with the solar plasma and how it influences the conversion of energy between its mechanical, magnetic, thermal, and radiative forms.
In the photosphere the thermal, kinetic, and magnetic energy have the same order of magnitude, so that this layer of the solar atmosphere plays
a key role because there the energy can easily be transformed from one form into another. This interaction is responsible for the creation of
a variety of magnetic structures, starting with sunspots ($\sim$10000-50000~km), pores ($\sim$1000-5000~km), and micropores ($\sim$300-1000~km) down to
bright points ($\sim$200~km or less), among the smallest known magnetic features.

Bright points (BPs) are brightness enhancements in the dark lanes separating solar granules. BPs are caused by magnetic flux concentrations
in the kilogauss range and are important for, e.g., the analysis of the solar influence on the Earth's climate. In the last 35 years, measurements
of satellite-based instruments revealed that the solar constant (which is the solar irradiance integrated over all wavelengths) is not constant, but
varies over an 11 year cycle by about 0.1\,\%. Surprisingly, the solar constant is on average higher during maximum solar activity than during minimum
(Willson \& Hudson 1988). The darkness of the sunspots is overcompensated by the increased brightness of BPs, whose number density is considerably higher
around sunspots than in the quiet Sun (Fr\"ohlich 2013; Solanki et al. 2013).

The magnitude of the variation of the solar irradiance is not uniformly distributed over the electromagnetic spectrum, but 60\,\% of the variations (Krivova et al. 2006) or
even more (Harder et al. 2009) are produced at wavelengths shorter than 400\,nm. A variation of the ultraviolet (UV) radiation changes the chemistry of the
stratosphere which can propagate into the troposphere and hence influence the weather and climate on Earth (London 1994; Haigh et al. 2010; Ermolli et al. 2012).
The most famous indication of a solar influence on the terrestrial climate is the so-called Maunder minimum, the period between 1645-1715 in which almost no
sunspots were observed and which coincides with the coldest part of the little ice age in Europe and Northern America.

One possibility to gain time series of the photosphere in the UV spectral range with high spatial and temporal resolution is to fly a telescope in the gondola
of a stratospheric balloon. Such balloon flights have a long tradition in solar physics, which was extended by the \sunrise{} mission described below
(an overview of the mission is given by Solanki et al. 2010 and Barthol et al. 2011).

\section{Mission and instrumentation}

\begin{figure*}[!ht] 
\begin{center}
\includegraphics[width=10cm]{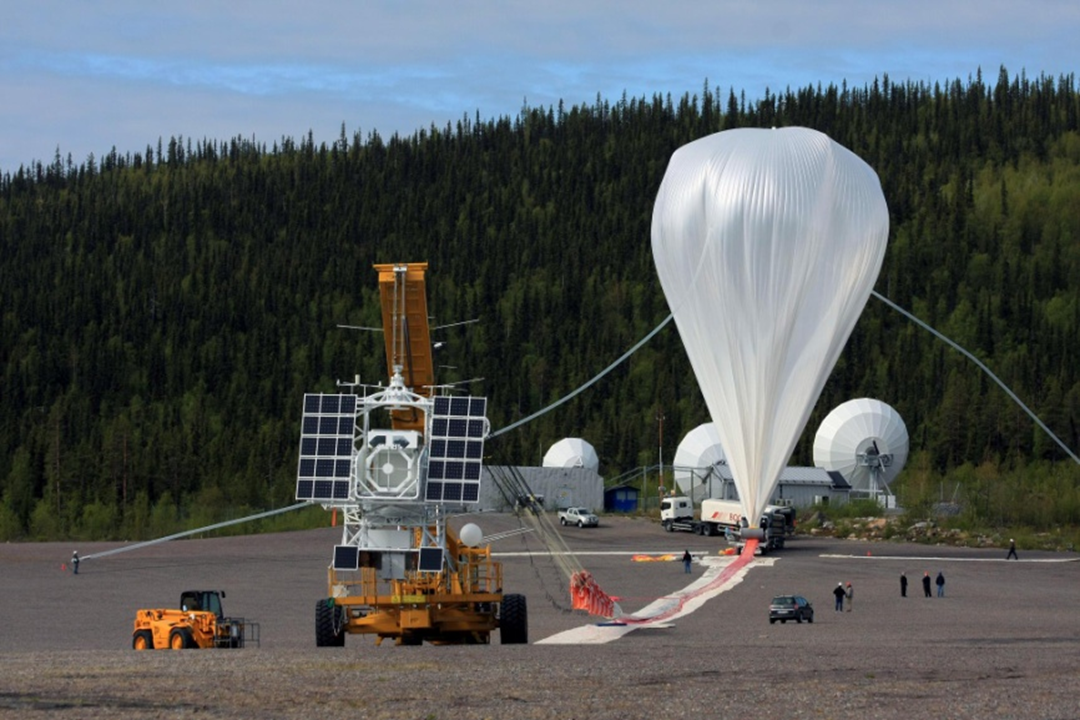}
\caption{The \sunrise{} observatory immediately before the launch of the first science flight.}
\label{fig1}
\end{center}
\end{figure*}

\sunrise{} is a balloon-borne solar observatory and consists of a 1\,m aperture Gregory telescope, a UV filter imager, a vector magnetograph for the visible
spectral range, and an image stabilization system. The 1\,m primary mirror makes \sunrise{} the largest solar telescope so far to leave the ground.
Up to now two science flights were carried out, the first one in June 2009 during minimum solar activity and a second one in June 2013 at a high activity level.
The two flights began from Esrange near Kiruna in northern Sweden (see Figure~\ref{fig1}). After a total flight time of about 6 days the gondola with the telescope
landed safely on a parachute in northern Canada so that the science data (which were stored on an onboard disk array) could be recovered. The mean flight altitude
of 37\,km allowed observations in the UV and visible spectral range that were free of atmospheric disturbances. Details about the telescope, the gondola's pointing
system, and the telemetry system can be found in Barthol et al. (2011). Berkefeld et al. (2011) describe the correlation tracker and the six-element Shack-Hartmann
wavefront sensor that controlled the telescope's focus mechanism and the tip/tilt mirror used for image stabilization, while Gandorfer et al. (2011) detail the
light distribution unit.

\subsection{SuFI}

The \sunrise{} Filter Imager (SuFI) is a phase-diversity assisted broadband imager for the near UV and violet spectral range. SuFI samples the solar photosphere
and chromosphere in five different wavelength ranges between 200\,nm and 400\,nm. The theoretical diffraction limit is 0\carcsec{}05, corresponding to 40\,km on
the solar surface. The phase-diversity prism in front of the $2048 \times 2048$ UV-enhanced CCD delivers a nominally focused image on one half of the detector,
while the other half receives an image with a defocus of one wave. In the post-processing, the two half images can be used for a removal of low-order aberrations.
The effective field of view (FOV) is $15\arcsec{} \times 40\arcsec{}$. The cadence of SuFI is up to one image per second, depending on the exposure time and the
observation mode. Detailed information on SuFI can be found in Gandorfer et al. (2011).

\subsection{IMaX}

The Imaging Magnetograph eXperiment (IMaX) is an imaging spectropolarimeter in dual-beam configuration that scans the highly Zeeman-sensitive Fe\,{\sc i} line
at 525.02\,nm. The cadence of IMaX is typically about 30\,s, the spectral resolution is 85\,m\AA{}, and the FOV is $50\arcsec{} \times 50\arcsec{}$. Several
operating modes are available. Most frequently used are a) the V5/6 mode in which the full Stokes vector is measured at five wavelength positions
(four within the spectral line and one in the continuum) with six accumulations to increase the signal-to-noise ratio to about $10^3$ in the continuum and b) the longitudinal
L12/2 mode in which only Stokes~$I$ and $V$ are measured at twelve wavelength positions with two accumulations. Phase-diversity information can also be obtained
since a plate can be inserted into the light path of one of the two synchronized $1024 \times 1024$ cameras. A full description of IMaX is provided by
Mart\'{\i}nez Pillet et al. (2011).

\section{Highlight results}

From the more than 40 refereed publications on science results obtained by the analysis of \sunrise{} data, a brief overview of a selection of these results
will be given in this section.

\subsection{Kilogauss fields resolved}

\begin{figure*}[!ht] 
\begin{center}
\includegraphics[width=8cm]{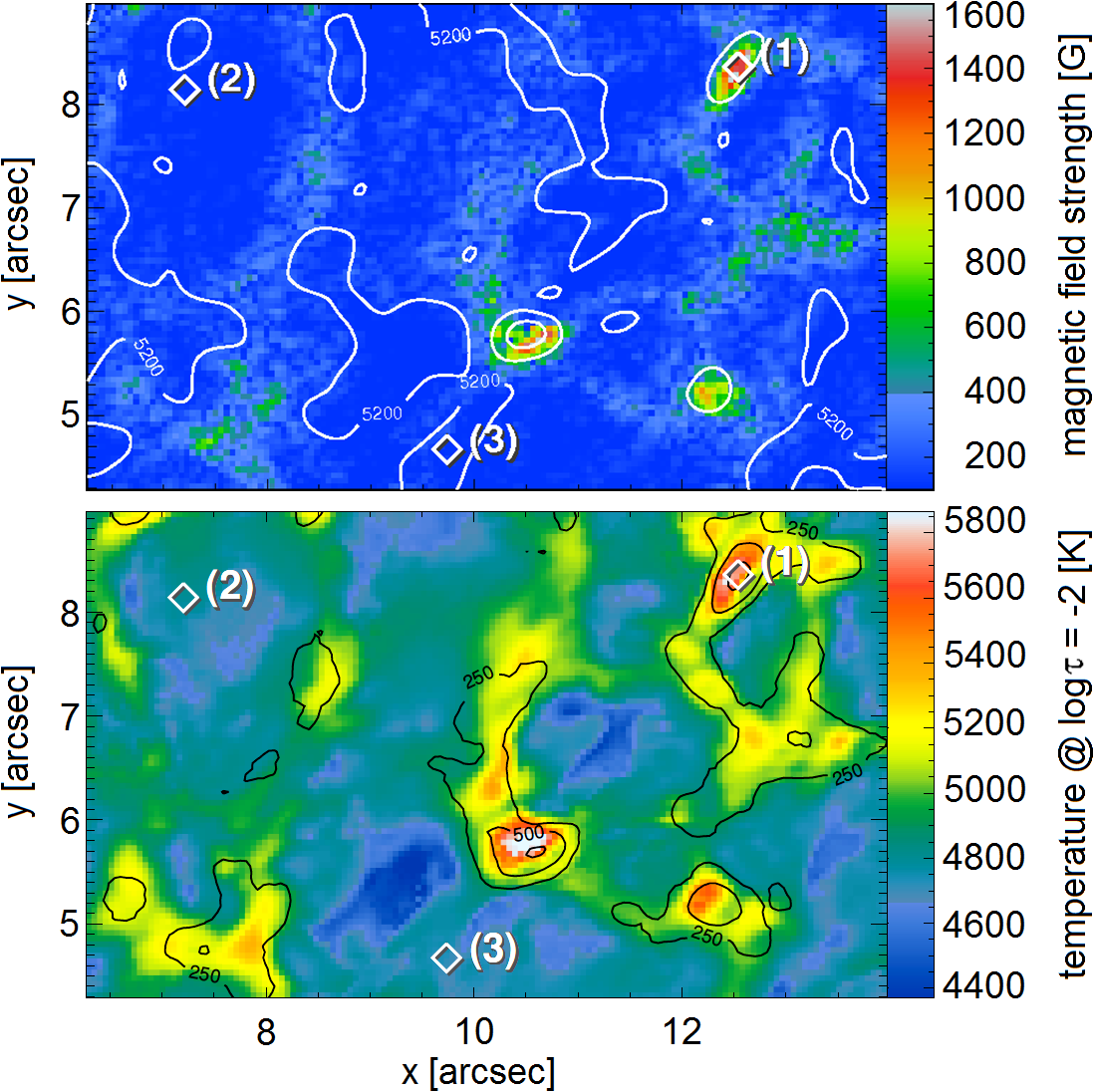}
\caption{Map of the magnetic field strength (top panel) and the temperature in the middle photosphere (bottom panel). The diamonds mark examples
of a magnetic patch (1), the inner part of a granule (2), and an intergranular lane (3). The white (black) contour lines show the temperature at $\log \tau = -1$
(magnetic field strength). Adapted from Lagg et al. (2010).}
\label{fig2}
\end{center}
\end{figure*}

Lagg et al. (2010) analyzed IMaX V5/6 data obtained from a quiet-Sun region at disk center. Stokes~$V$ maps displayed isolated flux patches.
The observed Stokes spectra were inverted with the SPINOR code (Frutiger et al. 2000) that numerically solved the radiative transfer equation. A simple
one-component model atmosphere with a height-dependent temperature and a height-independent magnetic field vector and line-of-sight velocity was used.
The retrieved magnetic field strength in the isolated flux patches is up to 1.5\,kG, while it is weak outside the patches (top panel of Figure~\ref{fig2}).
The flux patches' temperature in the mid photosphere is significantly enhanced with respect to its surroundings (bottom panel of Figure~\ref{fig2}), which
is consistent with semi-empirical flux-tube models. Since the kilogauss fields in the magnetic elements were retrieved without the introduction of a magnetic
filling factor we conclude that \sunrise{} could finally resolve the observed quiet-Sun flux tubes.

\subsection{Internal magnetic structure of a network element}

\begin{figure*}[!ht] 
\begin{center}
\includegraphics[width=12.5cm]{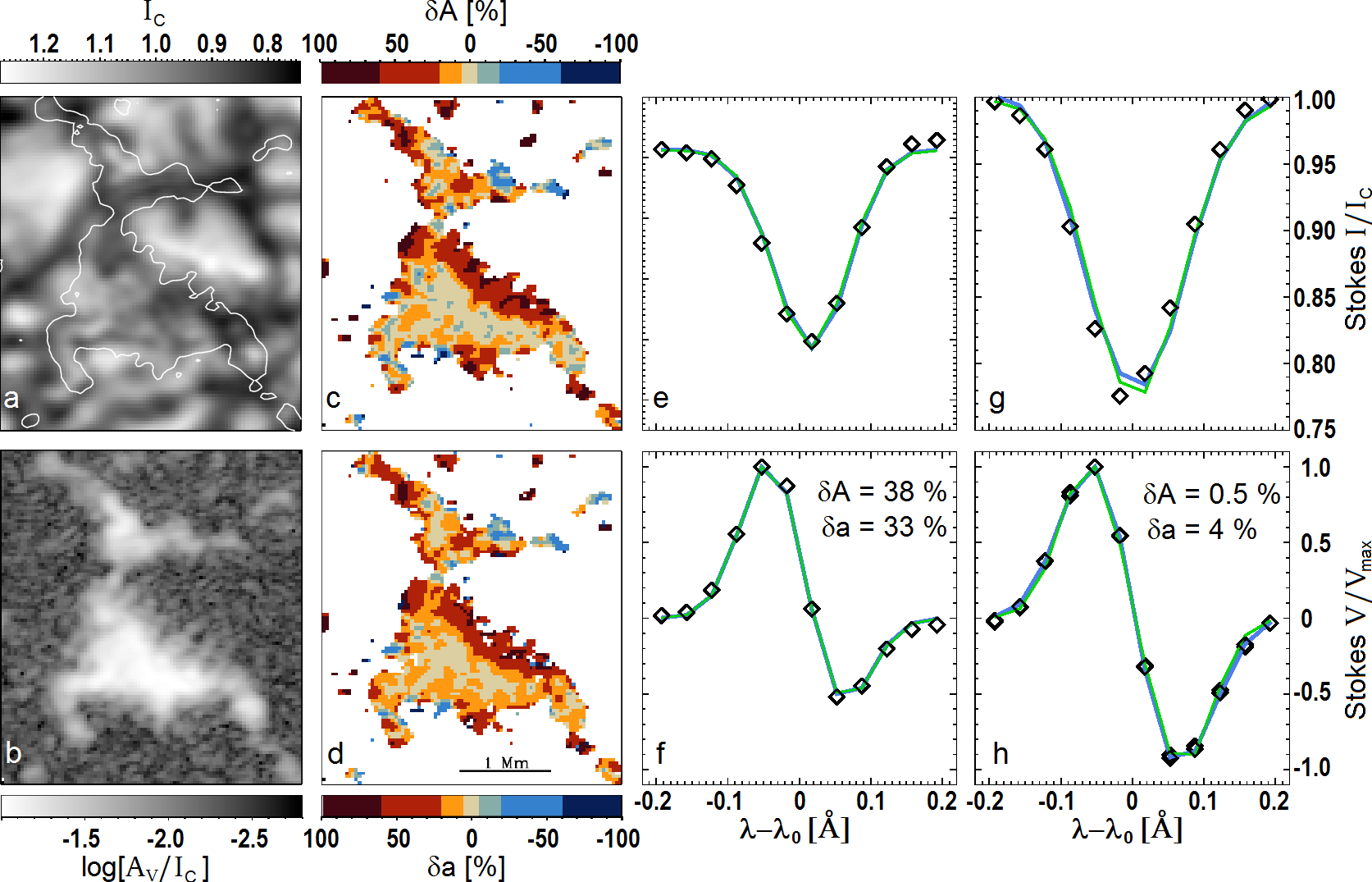}
\caption{Continuum intensity at 525.04\,nm (panel a), circular polarization amplitude (panel b), and Stokes~$V$ area (panel c) and amplitude asymmetry (panel d)
of the strongest flux region in the IMaX field of view. The white areas of panels c~\&~d harbor a signal-to-noise ratio of the Stokes~$V$ profiles lower than 5.
Panels e~\&~g (f~\&~h) display examples of Stokes~$I$ ($V$) profiles. Panels e~\&~f are created from a pixel near the border of the flux patch with a more
asymmetric $V$ profile, while panels g~\&~h represent the center of the flux region with relatively antisymmetric $V$ profiles. The diamonds exhibit the
observed profiles and the green and blue lines are best-fit results of various Stokes inversion models. Adapted from Mart\'{\i}nez~Gonz\'{a}lez et al. (2012).}
\label{fig3}
\end{center}
\end{figure*}

An investigation of the Stokes~$V$ asymmetries retrieved from IMaX L12/2 data showed that even the internal magnetic structure of a quiet-Sun network element could
be resolved with \sunrise{} (Mart\'{\i}nez~Gonz\'{a}lez et al. 2012). Both the amplitude and area asymmetry of the Stokes~$V$ profiles increase from very low values
at the core of the network element to values close to unity at its periphery (see Figure~\ref{fig3}). An asymmetry between the blue and red lobe of Stokes~$V$
profiles is caused by a gradient in magnetic field and velocity along the line of sight (LOS). The results of Mart\'{\i}nez~Gonz\'{a}lez et al. (2012) are consistent
with an expanding magnetic canopy, i.e. a region where the LOS passes through the magnetized atmosphere of the expanding flux tube in the upper photosphere,
then hits the nearly field-free atmosphere below the canopy before penetrating the $\tau = 1$ surface and entering the convection zone.

\subsection{Inclinations of quiet-Sun magnetic elements}

\begin{figure*}[!ht] 
\begin{center}
\includegraphics[width=8cm]{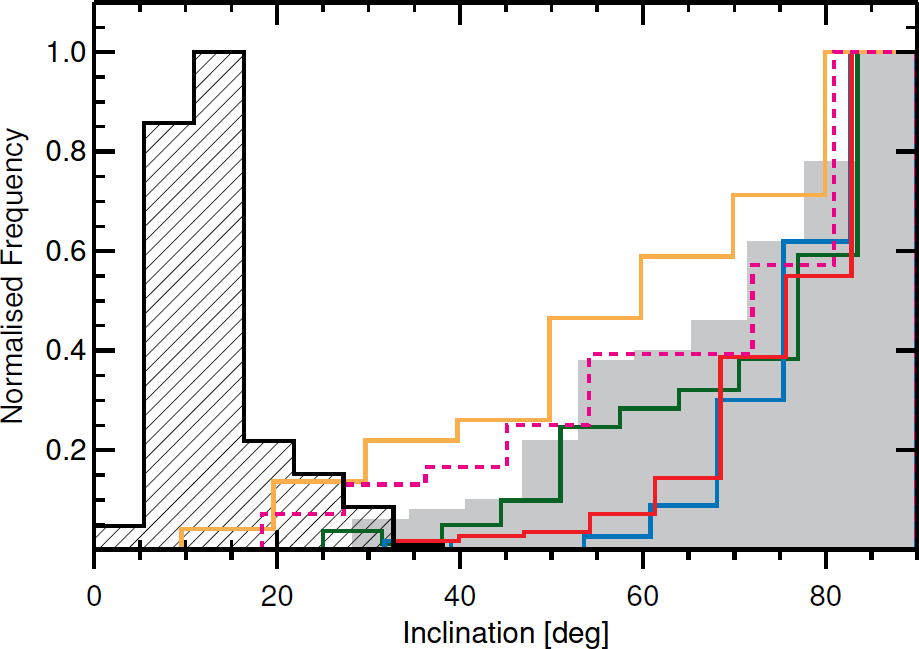}
\caption{Histograms of the bright magnetic element inclination relative to the line of sight. The black hashed histogram is obtained from a direct geometrical method.
The histograms on the right-hand side result from inverting Stokes profiles of the same set of magnetic features with various inversion codes applied to data treated in different ways.
From Jafarzadeh et al. (2014).}
\label{fig4}
\end{center}
\end{figure*}

The positions of a set of bright magnetic elements were determined by Jafarzadeh et al. (2014) from \sunrise{} images recorded in different spectral bands. An estimate of the formation height
of these spectral bands provided the inclinations of the magnetic elements from a direct geometrical method. This method returned that the magnetic field in BPs is nearly vertically oriented
with a narrow distribution. In contrast, the traditionally used inversions of Stokes profiles provided an almost horizontal field for the same set of magnetic elements (see Figure~\ref{fig4}).
Since the linear polarization signals in Stokes~$Q$ and $U$ are adversely affected by noise, the inversions overestimate the horizontal flux of such magnetic features by an order of magnitude.
The almost vertical orientations of the magnetic field in BPs were confirmed by a study of Riethm\"uller et al. (2014), who compared BP properties between \sunrise{} observations and magnetohydrodynamical
simulations after they were carefully degraded to take into account all the instrumental effects that were present during the observations (spatial and spectral resolution, stray light, noise).

\subsection{Bright point contrasts in the ultraviolet}

\begin{figure*}[!ht] 
\begin{center}
\includegraphics[width=8cm]{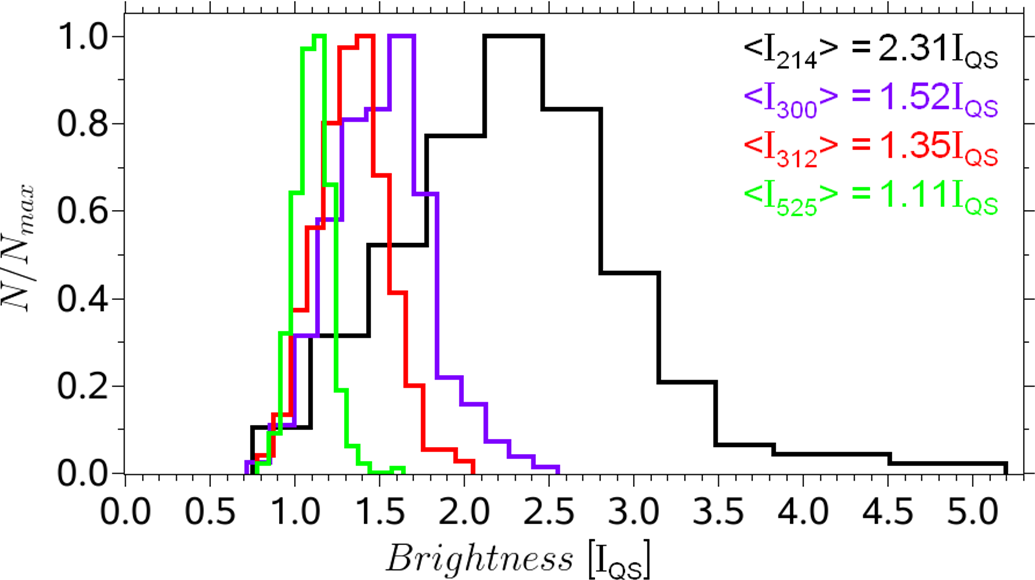}
\caption{Histograms of BP peak intensities at 214\,nm (black line), 300\,nm (magenta line), 312\,nm (red line), and 525\,nm (green line).
Adapted from Riethm\"uller et al. (2010).}
\label{fig5}
\end{center}
\end{figure*}

The UV capabilities of \sunrise{} allowed for the first time high-resolution observations of BPs in this important spectral range. Riethm\"uller et al. (2010)
analyzed quiet-Sun data at disk center and determined BP intensity contrasts that are considerable higher than the rms granulation contrast at all wavelengths
observed by SuFI. In particular at 214\,nm, BPs are up to 5 times brighter than the mean quiet-Sun, making them brighter than at any other spectral range studied
so far (see Figure~\ref{fig5}). The high BP contrasts and broad contrast distributions at short wavelengths confirm the assumption that BPs are important
for irradiance variations in the UV and hence for a possible solar influence on the climate on Earth.

\subsection{First high-resolution images of the Sun in the $279.6$\,nm M\lowercase{g}\,{\sc ii}\,\lowercase{k} line}

First analyses of data from the second \sunrise{} flight were carried out by Riethm\"uller et al. (2013) and by Danilovic et al. (2014). The first high-resolution images
of quiet Sun, active region plage, and sunspots in the Mg\,{\sc ii}\,k $279.6$\,nm line were presented and compared with images recorded in the core of the
Ca\,{\sc ii}\,H line. The Mg\,{\sc ii}\,k images show qualitatively very similar structures to those in Ca\,{\sc ii}\,H. In internetwork regions of the quiet Sun the
Mg\,{\sc ii}\,k images display reversed granulation or shock waves, while fibril structures are found in plage regions. The Mg filtergrams display higher intensity
contrasts and appear more smeared and smoothed in the quiet Sun than in the Ca channel. Both the Mg\,{\sc ii}\,k and the Ca\,{\sc ii}\,H line sample heights in the
chromosphere that are important to understand the heat transport from the photosphere to the corona.

\section{Conclusion}

Stratospheric balloon flights with the \sunrise{} observatory have provided high-quality UV and spectropolarimetric images with high spatial and temporal resolution
at different positions on the solar disk. The above examples have been listed to demonstrate the wide variety of studies carried out with \sunrise{} data. The low
solar activity level at the time of the first science flight means that mainly insights into the quiet Sun could be retrieved from these data. Since the
data analysis is still ongoing, in particular for the data of the second science flight at a much higher activity level, many more results are to be expected, e.g.,
on oscillations along Ca\,{\sc ii}\,H fibrils, on the relation between the magnetic flux density and intensity in the UV, on siphon flows in granule-size, twisted
magnetic loops, or on the power spectra of photospheric flows. A third flight of the \sunrise{} observatory is targeted for 2019 with fully renewed versions
of SuFI and IMaX in addition to a third scientific instrument, a slit spectrograph for scanning a photospheric and a chromospheric spectral line. This combination
of three instruments allows observations of the 3D structure of the magnetic field which is essential to answer the underlying physical questions.

\end{document}